\documentstyle[pre,aps,multicol,epsfig,tabularx]{revtex} 

\begin{document} 

\draft 

\title{Angle of repose and segregation in cohesive granular matter} 
\author{Azadeh Samadani and A. Kudrolli} 
\address{Department of Physics, Clark University, Worcester, MA 01610} 
 
\date{\today} \maketitle 

\begin{abstract} 

We study the effect of fluids on the angle of repose and the segregation of granular matter poured into a silo. The experiments are conducted in two regimes where: (i) the volume fraction of the fluid (liquid) is small and it forms liquid bridges between particles thus giving rise to cohesive forces, and (ii) the particles are completely immersed in the fluid. The data is obtained by imaging the pile formed inside a quasi-two dimensional silo through the transparent glass side walls and using color-coded particles. In the first series of experiments, the angle of repose is observed to increase sharply with the volume fraction of the fluid and then saturates at a value that depends on the size of the particles. We systematically study the effect of viscosity by using water-glycerol mixtures to vary it over at least three orders of magnitude while keeping the surface tension almost constant. Besides surface tension, the viscosity of the fluid is observed to have an effect on the angle of repose and the extent of segregation. In case of bidisperse particles, segregation is observed to decrease and finally saturate depending on the size ratio of the particles and the viscosity of the fluid. The sharp initial change and the subsequent saturation in the extent of segregation and angle of repose occurs over similar volume fraction of the fluid. Preferential clumping of small particles causes layering to occur when the size of the clumps of small particles exceeds the size of large particles. We calculate the azimuthal correlation function of particle density inside the pile to characterize the extent of layering. In the second series of experiments, particles are poured into a container filled with a fluid. Although the angle of repose is observed to be unchanged, segregation is observed to decrease with an increase in the viscosity of the fluid. The viscosity at which segregation decreases to zero depends on the size ratio of the particles. 
\end{abstract}

\pacs{PACS number(s): 45.70.Mg, 47.50.+d, 64.75.+g, 81.05.Rm} 

\begin{multicols}{2}  

\section {Introduction}
The presence of small amounts of liquid can have a considerable influence on the properties of granular matter. For example, the angle of repose of a wet granular pile is greater than that of a dry pile made of the same material. Cohesive forces are introduced because of liquid bridges that are formed between particles. Cohesivity causes jamming in the flow of granular matter even in a regime where dry granular matter may flow. Wet granular matter is also observed to segregate less than dry granular matter as particles cannot move easily relative to each other. Although these qualitative facts are well known, a detailed knowledge of the properties even in comparison with dry granular matter has not yet been attained. 

The properties of dry granular matter have attracted a considerable number of studies. Granular matter in a box which is tilted does not flow until a certain angle is exceeded at the surface. This angle usually called the maximum angle of stability $\theta_m$ depends on the frictional properties and the packing of the granular matter. However, if the granular matter is in motion, it can relax down to a lower angle which is called the angle of repose $\theta_r$. The existence of the lower $\theta_r$ may be due to the lower kinetic friction compared to static friction between the particles and other dynamical effects~\cite{jaeger90}. The observed hysteresis may also occur because there are infinitely many meta-stable states above $\theta_r$ in which the system can get trapped~\cite{mehta93}. 

In case of a very large pile of granular matter, it has been argued that $\theta_m$ is unaffected by the cohesivity introduced by the presence of the fluid. This is true provided that the frictional properties between the particles are unaltered due to the presence of a lubrication layer. The basic argument is that while frictional force which helps hold a pile together increases with the size of the pile, the cohesive force is constant and becomes insignificant in comparison~\cite{nedderman92}. However the effects can be important for a laboratory or industrial scale system. Recently, there have been a number of studies that have tried to relate the fluid content and its surface tension to observed increase in angles. $\theta_r$ measured by Schiffer and collaborators~\cite{hornbaker97,tegzes99} using the draining crater method is observed to initially increase linearly and then fluctuate around a constant value. For an ideal sphere-sphere contact, increasing the liquid content will lead to decreased cohesive force and hence a lowered $\theta_r$. Therefore they proposed that particle surface roughness plays an important role and a cone-plane type of contact between particles may be more appropriate. They also believed that the fluid coats the particles evenly. Mason {\em et al.}~\cite{mason99} have studied the distribution of the fluid on the surface of the particles using optical fluorescence microscopy and scanning electron microscopy. These studies appear to show that the fluid is in fact trapped in the menisci that are formed at the asperity on the grain surface. They have therefore claimed that the behavior of $\theta_m$ at low volume fractions of the fluid is consistent with a scaling theory based on the surface roughness of the grains~\cite{halsey98,bocquet98}. Other studies note that considerable differences between $\theta_r$ and $\theta_m$ can arise when humidity is present and have shown that the wetting properties of the fluid can be also important~\cite{fraysse99}. 

The identification of the occurance and the extent of segregation is of considerable importance in material handling in industry. Preferential percolation of small particles through layers of larger particles is usually identified as leading to segregation when granular flows occur at a free surface~\cite{bridgwater93,savage88}. Stratification can occur in such systems if rough and smooth shaped grains are present~\cite{makse97}. The presence of the fluid alters the percolation of particles~\cite{bridgwater78} and thus has a profound influence on the progress of segregation in granular matter. It is thus surprising that the effects of the fluid on the progress of segregation has not been investigated in detail till recently. In a recent publication, we reported that the presence of a fluid drastically reduces segregation in bidisperse granular matter that is poured into a silo~\cite{samadani00}. 

Although it is clear that the surface tension of the fluid is important in determining the cohesive force and hence $\theta_r$, the role of the viscosity of the fluid $\nu$ may be less obvious. It is a common observation that a finger dipped in honey feels sticky in comparison to a finger dipped in water. This occurs because stronger forces are required to move the liquid into regions that are vacated. Therefore viscous forces may have a significant effect on the dynamics of the grains. It is thus important to investigate the effect of $\nu$ on both $\theta_r$ and on size separation. Furthermore, viscous effects grow in importance for smaller dimensions. It may be possible to understand the effects of humidity on powders, that are smaller in size in comparison, by investigating effects of viscosity on segregation in granular systems. This may be useful because quantitative studies with powders are more difficult to conduct.

In this context, we report a detailed study of the angle of repose and segregation of granular matter in the presence of a fluid using high resolution digital imaging. Some of the results have been reported in a Letter~\cite{samadani00} and here we report additional data and analysis. We first investigate the effect of the size of the particles on $\theta_r$ as a function of the volume fraction of the fluid. We also discuss in detail our experimental observation of the effect of the viscosity $\nu$ of the fluid on $\theta_r$. We then study the extent of segregation by visualizing the color-coded glass particles using bidisperse glass beads with size ratio $r$. Interestingly, the decrease in segregation is observed to saturate at similar values of volume fraction as $\theta_r$. It can be also noted that small particles preferentially clump in a bidisperse mixture and this leads to subtle effects on the nature of the particle spatial-distribution as the volume fraction of the fluid is increased. As the volume fraction of the fluid increases and as the limit is approached where all the interstitial volume is saturated with the fluid, one can expect that the effects of the cohesion introduced by liquid bridges to become less important. This may have an effect on $\theta_r$ and size separation. However, there are practical difficulties in investigating high volume fractions because the fluid tends to drain. Therefore we also conducted experiments when the particles are completely immersed in a fluid to understand this limit. 

\section{Experimental Apparatus} 
A schematic of the experimental apparatus is shown in Fig.~\ref{exp_app}. A rectangular silo of dimensions $50.0\, {\rm cm} \times 30.5\, {\rm cm}$ and a width $w$ of $3.0\,{\rm cm}$ is used for the experiments.  The flow is visualized through the glass side-walls of the silo using a 1000 pixel $\times$ 1000 pixel Kodak ES 1.0 digital camera. The glass particles and fluids used are listed in Table~\ref{particles} and Table~\ref{fluids} respectively. The granular sample is prepared in batches by throughly mixing 1.0~kilogram of each of the two kinds of particles with the fluid before the mixture is placed inside the reservoir. The volume fraction $V_f$ of the fluid is calculated as the ratio of the volume of the fluid to the volume of all the particles. The volume of the particles corresponds to the weight of the particles divided by the density of the glass beads which is $2.4\,{\rm g cm}^{-3}$. 

The wet granular material is first filled into a reservoir and then drained through a pipe into the silo. In previous experiments we demonstrated that segregation can occur during draining of a wide silo~\cite{samadani99}, and therefore a tall cylindrical reservoir is used to minimize segregation from occuring during pouring. Pipes with various diameters are used to control the flow rate $Q$. Larger diameters are used at higher $V_f$ in order to maintain $Q \sim 2.2$~g s$^{-1}$. The reservoir is raised at a slow constant rate with a stepper motor and a system of pulleys. The slow upward velocity of the reservoir allows the particles to accumulate inside the pipe before flowing down the surface and reduces the kinetic energy of particles due to free fall. Such precautions are necessary otherwise particles acquire substantial kinetic energy during free fall and tend to bounce back from the surface several times leading to size separation and stratification in the case of bidisperse particles. Smaller particles bounce back higher from the surface of the pile and land further down the pile in contrast to what is usually observed, i.e. small particles at the top of the sandpile. This effect coupled with periodic avalanching observed at the surface can lead to alternate layers of mixed and small particles. By using the pipe, we ensure that the interaction of particles is restricted to the surface of the pile. Thus the number of mechanisms involved in segregation is reduced and the system is simpler.

We varied the time that the sample was mixed and found it had little effect on $\theta_r$. We also waited a varying amount of time before pouring to test the effect of evaporation. We found that only in the case of water, there was a small influence if we waited for one hour for volume fractions below $V_f < 2 \times 10^{-3}$. Therefore, to minimize the effects of evaporation, the mixture was poured immediately after mixing.

\section{The effect of interstitial fluid on the angle of repose}
A triangular shaped granular pile is formed after the material is poured into the silo. The flow of the granular material is continuous for dry particles, but becomes increasingly stick-slip as $V_f$ is increased. The flowing region is roughly 10 layers deep for dry particles, but the flow region becomes deeper and is not as well defined as $V_f$ is increased. The resulting pile for monodisperse glass particles is shown in Fig.~\ref{roughness} and it can be observed that the surface of the wet pile is at a greater inclination than the dry pile. 

\subsection{Size dependence of the angle of repose}
\begin{figure}
\begin{center} 
\centerline{\epsfig{file=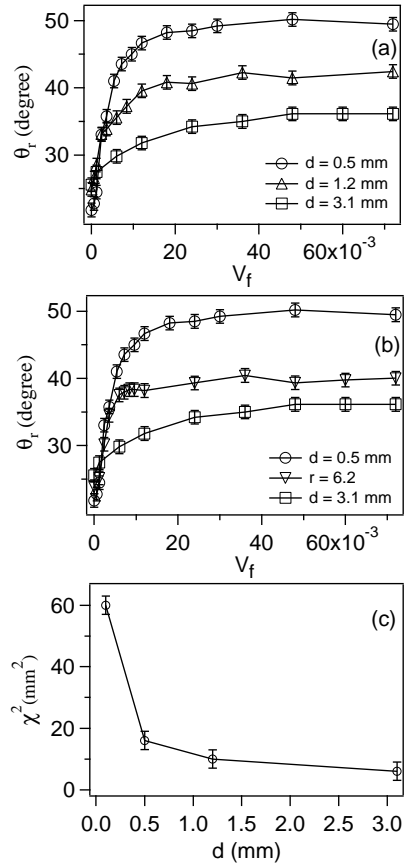,width=6.0 cm}} 
\end{center}
\caption{(a) Size dependence of the $\theta_r$ for $d = 0.5$ mm, $d = 1.2$ mm, and $d = 3.1$ mm glass beads. Note that $\theta_r$ is lower for larger $d$. (b) The saturation value of the $\theta_r$ of bimixtures are in between the $\theta_r$ of their components. (c) The mean square of the deviation of the surface from a straight line. The smaller particles are observed to stick together more readily, and make more clumps giving rise to a rougher surface and higher $\chi^2$.}
\label{AOR-sized} 
\end{figure}

$\theta_r$ is measured by obtaining images of the sand pile after the silo is filled and fitting a straight line to the surface of the pile. The obtained slope is averaged over 10-15 images as a centimeter deep layer of granular matter is added and a new surface is created.  $\theta_r$ is plotted as a function of $V_f$ for particles with various diameters $d$ in Fig.~\ref{AOR-sized}(a). $\theta_r$ is first observed to increase sharply with $V_f$ and then saturate approximately. The value of $V_f$ at which $\theta_r$ begins to saturate is $V_c$. The saturation value of $\theta_r$ is observed to increase with the size of the particles. This fact can be qualitatively noted from Fig.~\ref{roughness}. 

The increase and approximate saturation of $\theta_r$ is qualitatively similar to those earlier observed using the draining crater method~\cite{hornbaker97,tegzes99}. Quantitative differences exist perhaps because of the difference in the geometries. As discussed in the introduction, both the increase in $\theta_r$ and saturation is surprising if one assumes ideal spheres in contact. To explain this property, work has focused on the fact that particles are not perfectly smooth but have asperities. Using Mohr-Coloumb analysis and making assumptions for the height and distance over which the surface of the sphere fluctuates, Halsey and Levin~\cite{halsey98} found that $\theta_m$ can increase linearly and then saturate to a value given by:

\begin{equation}
\tan \theta_m \sim k + \frac{\sqrt{8} \pi k \Gamma}{d \rho_g g H} \sec[\tan^{-1}(k)]
\end{equation}
where, $k$ is coefficient of static friction, $\Gamma$ is the surface tension of the fluid, $g$ is acceleration due to gravity, $\rho_g$ is the density of the particles, and $H$ is height of the pile. It has been assumed here that the capillary force $F_{cap}$ between the particles is given by $\pi \Gamma d$. They assume that the saturation will occur approximately when the wetting is determined by the macroscopic curvature of the particles, which is greater than the scale of the roughness of the surface of the particles. According to the model, volume $V_c$ is given by $\lambda^2 d$ where $\lambda$ is length scale over which fluctuations occur. The model predicts that $\theta_m$ decreases with $d$, which is consistent with our observations. Although there is a qualitative agreement between our data for $\theta_r$ and the model for $\theta_m$~\cite{halsey98}, there is considerable disagreement related to the value of $V_c$. For example for 1 mm particles,  where the particle surface fluctuations are about $1 \mu$m, the model predicts saturation for $V_c \sim 10^{-5}$. However the observed $V_c$ is significantly higher and is approximately $5 \times 10^{-3}$. 

While it is possible that the disagreement can arise because $\theta_m$ is calculated and $\theta_r$ is measured, additional effects may have to be taken into account in calculating the average cohesive force due to capillary bridges. For example, it has been assumed that only the capillary forces between particles in contact is important. However it is possible that as $V_f$ increases, particles that are slightly away from each other can also form liquid bridges. Therefore, it is possible that the average number of liquid bridges per particle increases with $V_f$. This effect may be important in determining the regime over which $\theta_r$ can increase. 

We plot in Fig.~\ref{AOR-sized}(b), the $\theta_r$ for particles with $d = 0.5$ mm and $d = 3.1$ mm and the bidisperse mixture (BD-3) to illustrate the behavior of $\theta_r$ when particles of different diameters are present. The $\theta_r$ for the bidisperse mixtures are located in between the $\theta_r$ of their components. A similar effect is observed for the other size ratios. 

Besides the increase of $\theta_r$ due to the addition of the fluid, the surface of the pile is observed to become rough [see Fig.~\ref{roughness}(b),(d)]. In addition, the roughness of the surface of the pile is greater for the smaller particles. We plot the mean square of the deviation of the surface from a straight line ${\chi}^2$ for the particles mixed with water for a fixed $V_f = 24 \times 10^{-3}$ in Fig.~\ref{AOR-sized}(c) to quantitatively show the increase in roughness with decreases in particle size. We observe that smaller particles clump more than larger particles at similar $V_f$. We therefore believe that the increased roughness of the surface is because of clumping of the particles. As the particles form more clumps, the surface becomes increasingly rough and ${\chi}^2$ grows. The standard deviation of $\theta_r$ obtained by repeating measurements for a particular $V_f$ was reported in Ref.~\cite{tegzes99} as a measure of clumping. The reported observations are consistent with our more direct measurement of the surface heights. We will later see that the preferential clumping of small particles has an important effect on the progress of segregation and introduces layering.

\subsection{Effect of viscosity of the fluid on the angle of repose.}
To investigate the effect of viscosity on the flow and the angle of repose, we kept the surface tension of the fluid constant and change $\nu$ of the fluid. Because water and glycerol have very similar surface tension, we used mixtures of these two fluids to change $\nu$~\cite{viscosity}. Glycerol dissolves in water and the mixture is homogenous after it is stirred for a few minutes. Using water-glycerol mixtures we obtained $\theta_r$ as a function of $V_f$.  

\begin{figure}
\begin{center} 
\centerline{\epsfig{file=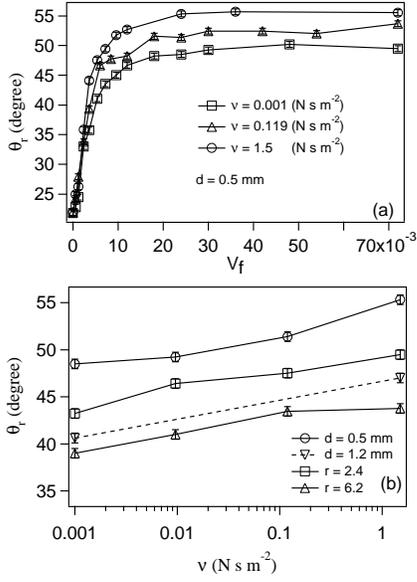,width=6.0 cm}} 
\end{center}
\caption{(a) $\theta_r$ for monodisperse particles as a function of $V_f$, for water ($\nu = 10^{-3}$ N s m$^{-2}$), glycerol ($\nu = 1.5$ N s m$^{-2}$) and water-glycerol mixture ($\nu = 0.12$ N s m$^{-2}$). (b) $\theta_r$ as a function of $\nu$ for different bimixture and monodisperse particles for $V_f = 24 \times 10^{-3}$.}
\label{aor-v} 
\end{figure}

In Fig.~\ref{aor-v}(a) we plot $\theta_r$ for $d = 0.5$ mm beads as a function of $V_f$, for water, glycerol, and a water-glycerol mixture. Fig.~\ref{aor-v}(a) shows that $\theta_r$ increases with $\nu$. The increase is observed to be sharper and the saturation is observed to occur at higher values for glycerol compared to water. We chose $V_f = 24 \times 10^{-3}$ in the saturated regime to plot $\theta_r$ as a function of $\nu$ for monodisperse particles and bidisperse mixtures in Fig.~\ref{aor-v}(b). We observe that saturation value for $\theta_r$ depends on $\nu$ and increases with $\nu$. 

We have measured $\theta_r$ up to 48 hours after the granular mixture is poured into the silo. While a few local rearrangements occur up to a few minutes after pouring, we do not find measurable changes in the surface, thus indicating the absence of creep. We have thus shown that the viscosity of the fluid clearly influences the observed $\theta_r$. In the next section we will discuss the viscous force between the particles due to the fluid.

\subsection{Estimates of the viscous force}

Let us consider two particles that are moving away from each other. A schematic diagram is shown in Fig.~\ref{force_schematic}(a). Fig.~\ref{force_schematic}(b) shows 0.1 mm particles mixed with glycerol for $V_f = 24 \times 10^{-3}$. Based on Reynolds lubrication theory, an expression for the viscous force has been derived for two identical rigid spherical surfaces~\cite{pitois00,Xu01,persson98}. The theory relates the pressure $P$ generated in the liquid to the relative displacement of the two particles as: 
\begin{equation}
\frac{d}{dr_1}(r_1 H^3(r_1) \frac{dP(r_1)}{dr_1}) = 12\nu r_1 \frac{dh}{dt},
\end{equation}
where, $H(r_1) = h_0 + 2r_1^2/d$ is the distance between the two surfaces at radial distance $r_1$ from the center [see Fig.~\ref{force_schematic}(a)]. If the particles are completely immersed in the fluid, the expression for the viscous force is:
\begin{equation}
F_{vis} = -\frac{3}{8}\pi \nu d^2 \frac{1}{h_0} \frac{\partial h_0}{\partial t}.
\end{equation}
For a liquid bridge, there is a correction coefficient~\cite{pitois00} to Eq. (3) and the expression for the force is:
\begin{equation}
F_{vis} = -\frac{3}{8}\pi \nu d^2 (1- \frac{h_0}{H(R)})^2\frac{1}{h_0} \frac{\partial h_0}{\partial t},
\end{equation}
where, $R$ is the radius of the contact area. This equation can be related to the volume of the liquid bridge using $V = \int_0^{R}2 \pi r_1 H(r_1)dr_1$. Pitois {\em et. al.}~\cite{pitois00} experimentally investigated the effect of viscosity of the fluid on the properties of a liquid bridge between two moving spheres. They showed that Eq.(4) fits their experimental data.

The value of the viscous force in our experiments can be estimated assuming that $\frac{\partial h_0}{\partial t}$ is of the order of the velocity of the particles flowing down the surface. For 1 mm particles, if $h_0$ is of the order of 10 $\mu$m then $F_{vis} \sim 10^{-7}$ N for water and $F_{vis} \sim 2 \times 10^{-4}$ N for glycerol. The capillary force estimated using $F_{cap} = \pi \Gamma d$ for 1 mm particles is $10^{-4}$ N. Therefore the viscous force for glycerol is relevant whereas for water the viscous force may be insignificant. These improved results are also consistent with our previous estimates obtained neglecting the curvature of the particles~\cite{samadani00}.

Because viscous force decreases with relative velocity it appears surprising that the viscosity of fluid plays any role at all in determining $\theta_r$ in our experiments and we return to this point later in the paper. However it is clear that such forces are important in determining the extent of segregation as it occurs when the particles are in motion. 

\section{Effect of interstitial fluid on size segregation}
To study the segregation of the particles in the pile, we use different colors for the two kinds of particles. Fig.~\ref{seg-mix}(a) shows a pile after dry bidisperse granular material has been poured inside the silo. Here the small particles appear white and the large particles appear black. Thus strong size separation is observed as the granular matter flow down the inclined surface. Two main mechanisms are important in determining the observed spatial distribution: (i) the void filling mechanism where smaller particles percolate through the larger particles and are thus found at the bottom of the flow~\cite{savage88}, and (ii) the capture mechanism where the smaller particles which are more sensitive to surface fluctuations are stopped at the top of the pile before the larger particles~\cite{boutreux96}. Segregation is observed to vanish when a small amount of fluid is added. Fig.~\ref{seg-mix}(b) shows the situation where $V_f = 6 \times 10^{-3}$, which corresponds to {\em less than 1\% by volume} of water in the system. Thus a very small amount of fluid is observed to prevent segregation from occuring.

To parameterize the extent of segregation, a histogram of the ratios of the two types of particles in a $3.5\,{\rm mm}^2$ area is measured using the light intensity. The light intensity is a monotonic function of the density ratio of the particles and is measured by using pre-determined weight ratios of particles in a separate series of calibration experiments. The histogram of local particle density ratio $P_w$ of the two kinds of particles in the pile is thus obtained and is plotted in Fig.~\ref{seg-mix}(c),(d). The peaks in distribution can be fitted to a Gaussian and the mean value is used to determine the most common particle ratios $a$ and $b$.  The segregation parameter is defined as $s= \frac{(b-a)}{100}$ which has a value between 0 and 1 and describes the extent of segregation.

\subsection{Dependence of the segregation on the size ratio}
\begin{figure}
\begin{center} 
\centerline{\epsfig{file=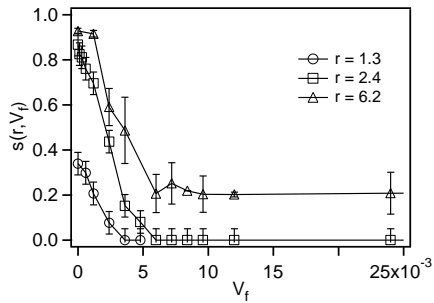,width=6.0 cm}} 
\end{center}
\caption{The extent of segregation $s$ observed in bidisperse particles as a function of $V_f$ of water for various size ratios $r$. Segregation decreases and saturates, when $V_f$ is greater than $V_c$. Segregation can persist if $r$ is large.}
\label{size-dependence} 
\end{figure}

$s$ for three different bidisperse granular mixtures BD-1, BD-2, BD-3 (see Table~\ref{particles}) are plotted in Fig.~\ref{size-dependence}(a) as a function of $V_f$ of water. $s$ depends on the size ratio $r$ of particles, and goes to zero at lower $V_f$ for smaller $r$. However, segregation can persist even in the presence of fluid although the extent is considerably smaller. A phase diagram of the observed segregation as a function of $r$ and $V_f$ can be found in Ref.~\cite{samadani00}.

\subsection{Discussion of segregation on $r$ size ratio}

The observed properties can be described in terms of the differences in the strength of interactions between particles due to difference in the sizes. In the presence of a wetting fluid, there are three possible interaction between the particles: interactions between the (i) small-small particles $S-S$, (ii) small-large particles $S-L$, and (iii) large-large particles $L-L$. In addition, the particles and clumps will also interact with the inclined surface on which they flow. 

At small $V_f$ the fluid is observed to selectively coat the small particles forming clumps of small particles due to the cohesive forces discussed earlier. Most of these interactions are in form of $S-S$. There is less percolation of small particles through larger particles thus reducing $s$. By increasing the amount of fluid further, the size of clumps increases, the interaction of $S-L$ also become important, resulting in even smaller $s$. Finally,  above a certain $V_f \sim V_c$ all types of interaction can be present, clumps are formed with both kinds of particles and segregation saturates. For large $r$ clumps of $S-L$ appear but are not strong enough and segregation decreases but saturates to non-zero values. 

\subsection{Layering instability}
\begin{figure}
\begin{center} 
\centerline{\epsfig{file=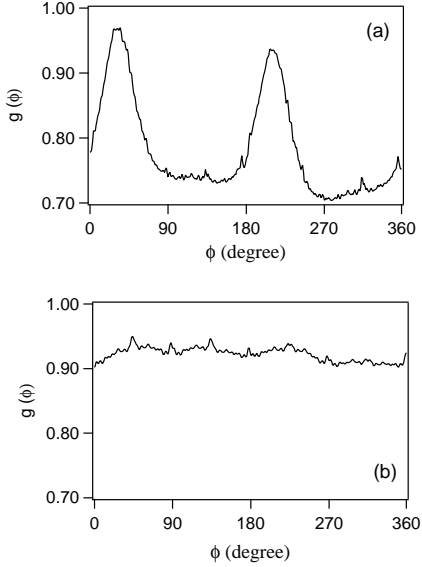,width=6.0 cm}} 
\end{center}
\caption{Azimuthal correlation function of the density of the particles for (a) $V_f = 1.2 \times 10^{-3}$ and (b) $V_f = 6 \times 10^{-3}$. The correlation function show two distinct peaks at the angles parallel to the surface. The peaks fade for higher $V_f$. The strongest layering is observed for $r = 2.4$.}
\label{correlation} 
\end{figure}

Fig.~\ref{layering}(a) shows the intermediate situation where partial segregation is observed. We obtain the azimuthal correlation function $g(\phi)$ of the particle density inside the pile to characterize the layering where $g(\phi)$ is given by:
\begin{equation}
g(\phi) = \sum_{x,y,r_2} \frac{I(x,y) \times I(x+ r_2 \cos(\phi),y + r_2 \sin(\phi))}{I(x,y)^2}
\end{equation}
where, $I(x,y)$ is the particle ratio at position $(x,y)$ and $r_2$ and $\phi$ define the distance from the point where correlation is calculated. The correlation function was calculated for a square region in the middle of the pile. The peaks were observed to decrease in amplitude and are hidden by the noise if the whole pile was used for averaging. Fig.~\ref{correlation}(a) shows $g(\phi)$ for the image shown in Fig.~\ref{layering}(a). The correlation function has two distinct peaks corresponding to the angles parallel to the surface.  The peaks are observed to decrease in amplitude as layering fades with increased $V_f$. Finally for the complete mixture we observe the flat line shown in Fig.~\ref{correlation}(b) for $V_f = 6 \times 10^{-3}$. The correlation function does not show peaks for larger size ratios consistent with the visual observation. 

Although the layering is not as periodic as in {\em dry} mixtures of rough and smooth particles~\cite{makse97}, some stratification is clearly observed below $V_c$. We believe that the layering is related to the selective coating and clumping of smaller particles (see Fig.~\ref{layering}). A clump of small particles effectively behaves as a particle with a rough surface which can be greater than the individual large particle. Makse {\em et al.} have shown that stratification can occur when two species with different surface roughness are present in granular flows when the size of the rough particles exceeds the smoother particles~\cite{makse97}. The features observed in Fig.~\ref{layering} appear to be related to this mechanism, although the increased stick-slip nature of the flow at the surface brought about by the addition of the fluid makes the layering aperiodic.  The strongest layering is observed for BD-2 and decreases at higher $r$ where the clumps of small particles stay smaller than the larger beads. At higher $V_f$, the average size of clumps increases and the interaction between small and large particles also become important. Thus clumps with small and large particles and layering disappears.

\subsection{Viscosity dependence of the segregation}
\begin{figure}
\begin{center} 
\centerline{\epsfig{file=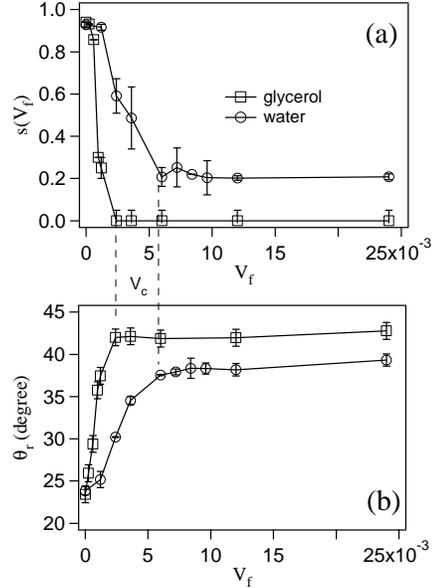,width=6.0 cm}} 
\end{center}
\caption{Viscosity dependence of the $s$ and $\theta_r$ for $r = 6.2$ as a function of $V_f$. Note that the saturation value $V_c$ is similar for $s$ and $\theta$.}
\label{viscosity-d} 
\end{figure}

Next we examine the effect of the viscosity of the fluid on the extent of segregation. As in the measurements for $\theta_r$, we change the viscosity of the fluid by using mixtures of water and glycerol to keep the surface tension almost constant. $s$ corresponding to water and glycerol is plotted in Fig.~\ref{viscosity-d}. The segregation is observed to decrease to zero at a lower value of $V_f$ in the case of higher glycerol, which has $\nu$ a thousand times that of water. In fact, although some incipient amount of segregation is observed for large $r$ in the case of water, segregation is observed to completely dissappear in the case of glycerol for all measured $r$. 

In Section III.C we discussed the strength of the viscous forces that may be important in determining percolation. For the particles ($d \sim 1.2$ mm) used in the experiments, and for glycerol or water as interstitial fluid, the saturation force due to surface tension is of the order of $10^{-4}$ N.  Therefore the viscous force is greater than the capillary force in the case of glycerol. The viscous force damps velocity fluctuation as it increases with relative velocity between particles. Because velocity fluctuations and percolation are required for segregation, $s$ is therefore observed to be lower at higher $\nu$ for similar $\Gamma$. Since the linear increase of $\theta_r$, and decrease of $s$ occurs below a similar $V_c$, and both of these quantities saturate above this volume fraction, the cohesion due to the fluid appears to have the same effect on both $\theta_r$ and $s$.

\section{Experiments when particles are immersed in the fluid}
In this section, we study the angle of repose and the progress of segregation when the bidisperse glass particles are poured into a silo filled with various liquids with different $\nu$. The procedure and the system is similar to the previous sections except that in these experiments, we first fill the silo with the fluid and then pour the dry granular matter. In this case only viscous forces are present and capillary forces are not present because liquid bridges are absent. 

Because the terminal velocity for particles decreases with $\nu$, they take a significantly longer time to drain in a viscous fluid compared to that in air and $Q < 0.02$ g s$^{-1}$ in glycerol. As $\nu$ is increased, the particles are increasingly observed to deviate from a downward trajectory and are deflected further down the slope even before coming in contact with the surface. The reason for the deflection of the particles can be qualitatively understood as follows. For a particle moving with velocity $v$, a boundary layer develops in the fluid which is proportional to $\sqrt{\nu d/\rho v}$. This estimate is essentially from dimensional arguments~\cite{batchelor70}. As the distance between the particle and the surface becomes comparable to the size of this boundary layer, the particles feel a net horizontal force similar to the viscous forces discussed earlier due to fluid expulsion. These forces tend to deflect the particles further down the slope. 

We observed $\theta_r$ is $\sim 24^0 \pm 1^0$ for mono and bidisperse glass particles for not only water and glycerol but also using polybutene (see Table~\ref{fluids}). Unlike the case of partial $V_f$, no systematic dependence of $\theta_r$ is observed on the viscosity. Furthermore, the observed $\theta_r$ in the fluid is very similar to that observed in air. This appears to suggest that the viscous forces discussed in Section III.C may not be relevant in the absence of liquid bridges. Although it must be noted that the experiments conducted in the regime where the particles are mixed with a small volume fraction of fluid is somewhat different in nature than these experiments because of the presence of the boundary layer. 

Experimental work investigating liquid bridges between moving spheres~\cite{pitois00,gaudet96,mckinley00} has shown that bridge rupture distance and time between particles increases with viscosity. It is therefore possible that the increases in $\theta_r$ with viscosity in the case of partial $V_f$ is related to the increase in the number and change in structure of liquid bridges with viscosity.

\begin{figure}
\begin{center} 
\centerline{\epsfig{file=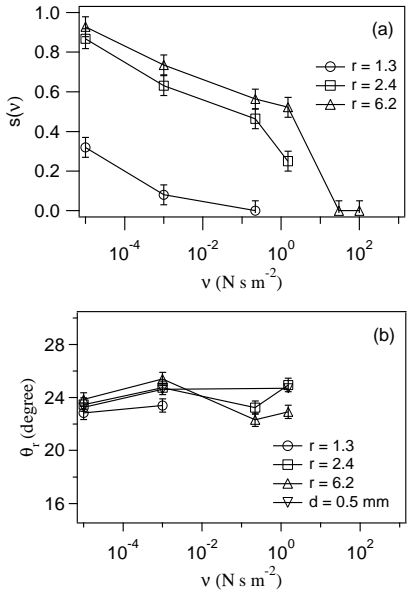,width=6.0 cm}} 
\end{center}
\caption{(a) The extent of segregation $s(\nu)$ is observed to decrease as a function of $\nu$ of the fluid. (b) $\theta_r$ as a function $\nu$ is approximately constant. This is in contrast to observations for partial $V_f$ where $\theta_r$ is observed to increase.}
\label{viscosity} 
\end{figure}

Fig.~\ref{immersed}(a) shows an image of the granular pile after bidisperse particles (BD-2) were poured into the silo filled with glycerol. Separate series of experiments to calibrate the intensity of light scattered by particles to the number of particles were first conducted when particles were completely immersed in the various fluids. Using the same procedure as before, we measured the histogram of light intensity to determine $s$ [see Fig.~\ref{immersed}(b)]. Comparing Fig.~\ref{immersed}(a) to Fig.~\ref{seg-mix}(a) shows a significant drop in $s$, when particles are poured in glycerol instead of air. The $s$ is plotted in Fig.~\ref{viscosity}(a) as a function of $\nu$ for three different $r$. The segregation is observed to decrease as a function of $\nu$ and drops to zero at high $\nu$ which depends on $r$. 

The observed decreases in segregation with $\nu$ can be explained as follows. The boundary layer washes away the details of the surface roughness of the pile at higher $\nu$. Thus the capture mechanism which is sensitive to surface roughness decreases in importance with $\nu$. Furthermore, velocity fluctuations at higher $\nu$ are damped out as particles reach terminal velocity over a short distance leading to a reduction in percolation of particles~\cite{bridgwater78}. Therefore the segregation decreases because the two mechanism which cause segregation in dry granular matter diminish in strength. 

It is well known that sedimentation of different sized particles (for example in lakes) cause layering of different size particles to occur at the bottom because the terminal velocity reached by a spherical particle is proportional to $\sqrt{d}$.  This does not occur in our geometry because particles are constantly fed into the system from the reservoir.

\section{Conclusions}

In conclusion, the role of the size ratio of the particles, volume fraction, surface tension and viscosity of the fluid on the extent of segregation and the angle of repose of a granular pile is clarified based on experiments and physical arguments. The angle of repose of the granular matter is observed to increase sharply as the volume fraction of the fluid is increased and then saturates. The saturation occurs at a higher $V_f$ than estimated by using only particle roughness ideas. We observe that the viscosity of the fluid has a significant effect on the angle of repose of the pile. A sharp reduction of segregation is observed in the granular flow when a small volume fraction of fluid is added. The sharp changes in the angle of repose and segregation occurs over similar volume fractions suggesting that cohesivity has the same effect on both properties. The experiments point to a need for examining the role of the number of liquid bridges between particles in determining the angle of repose. Our experiments appear to indicate that the changes in {\em the number of liquid bridges} and their structure with volume fraction, surface tension and viscosity of the fluid may be very important in determining the properties of wet granular matter.

We thank Jacob Goodman for help in acquiring data, and G. McKinley for bringing Ref.~\cite{pitois00} to our attention. This work was supported by the National Science Foundation under grant number DMR-9983659, the Alfred P. Sloan Foundation, and by the donors of the Petroleum Research Fund.



\begin{figure}
\begin{center} 
\end{center}
\caption{Schematic diagram of the experimental setup. The reservoir containing the initial mixture is pulled up using a stepper motor in order to keep the height through which the mixture falls constant.}
\label{exp_app} 
\end{figure} 


\begin{figure}
\begin{center} 
\end{center}
\caption{Images of a pile of dry and wet monodisperse particles. (a) $V_f = 0$, $d = 0.5$ mm, (b) $V_f = 10^{-2}$, $d = 0.5$ mm, (c) $V_f = 0$, $d = 3.1$ mm, (d) $V_f = 10^{-2}$, $d = 3.1$ mm. $\theta_r$ is observed to be greater for the wet case and the surface of the pile of small particles is observed to be rough compared to the larger particles.}
\label{roughness} 
\end{figure} 


\begin{figure}
\begin{center} 
\end{center}
\caption{(a) Schematic of a liquid bridge between two particles. (b) Image of a liquid bridge between two 0.1 mm particles corresponding to $V_f = 24 \times 10^{-3}$.}
\label{force_schematic} 
\end{figure}


\begin{figure}
\begin{center} 
\end{center}
\caption{Image of the granular pile after bidisperse glass beads are poured into the silo ($r = 2.4$). The small particles are white and the large particles are black. (a) $V_f = 0$, (b) $V_f = 6 \times 10^{-3}$. Strong segregation is observed for the dry case, the segregation is drastically reduced when less than 1\% of water is present in the mixture. (c)-(d) The histogram of local particle density ratio of the two kinds of particles in the pile. The peaks in the distribution can be fitted to Gaussian and the mean value used to determine the most common particle ratios.}
\label{seg-mix} 
\end{figure} 


\begin{figure}
\begin{center} 
\end{center}
\caption{(a) Strong layering is observed at $V_f$ below $V_c$ for $r = 2.4$ (here $V_f = 1.2 \times 10^{-3}$). (b) The fluid is observed to selectively coat the smaller particles, resulting in clumps of small particles. In this case the clumps of small particles are larger than the larger particles.}
\label{layering} 
\end{figure}


\begin{figure}
\begin{center} 
\end{center}
\caption{Image of the granular pile after bidisperse glass beads ($r = 2.4$) are poured into a silo filled with glycerol. Compare the extent of segregation to Fig.~\ref{seg-mix}(a), where air is the interstitial medium. The segregation is drastically reduced when particles are poured into a more viscous medium. (b) The histogram of local particle density ratio of the two kinds of particles in the pile.}
\label{immersed} 
\end{figure}


\begin{table}\begin{center} \begin{tabular}{l c c c c c} \hline  
Monodisperse \, & $d$(mm) \,\, &\,\,\,\, $\theta_r$\, \\ \hline \hline 
mono-1&0.1 $\pm$ 0.1\,\,\,\,\,\,\,\,  &$23.5^\circ\pm 0.5^\circ$\\

mono-2&0.5 $\pm$ 0.1\,\,\,\,\,\,\,\, &$24.0^\circ\pm 0.5^\circ$\\

mono-3&0.9 $\pm$ 0.1\,\,\,\,\,\,\,\,&$24.0^\circ\pm 0.5^\circ$\\

mono-4&1.2 $\pm$ 0.1\,\,\,\,\,\,\,\,&$24.0^\circ\pm 0.5^\circ$\\

mono-5&3.1 $\pm$ 0.1\,\,\,\,\,\,\,\, &$25.5^\circ\pm 0.5^\circ$\\\hline
\end{tabular} \end{center} 

\begin{table}\begin{center} \begin{tabular}{l c c c c} \hline 
Bidisperse \, & $d$ large \,\, & $d$ small &  $r$ \,\,\,\,\,& $\theta_r$\, \\ \hline \hline 

BD-1  &mono-4\,\,\,\,\,\,&mono-3\,\,\,\,\,\,&1.3\,\,\,\,\,\,&$23^\circ\pm 0.5^\circ$ \\

BD-2  &mono-4\,\,\,\,\,\,&mono-2\,\,\,\,\,\,&2.4\,\,\,\,\,\, &$23^\circ\pm 0.5^\circ$ \\ 

BD-3  &mono-5\,\,\,\,\,\,&mono-2\,\,\,\,\,\,&6.2\,\,\,\,\,\,&$25^\circ\pm 0.5^\circ$\\

BD-4  &mono-5\,\,\,\,\,\,&mono-1\,\,\,\,\,\,&31\,\,\,\,\,\, &$24^\circ\pm 0.5^\circ$\\
\hline \end{tabular} \end{center} 
\caption{Monodisperse and bidisperse mixtures of spherical glass particles used in the experiment. $d$ is the diameter of particles, $r$ the size ratio, and $\theta_r$ the angle of repose of the dry particles.}
\label{particles} 
\end{table}

\begin{center} \begin{tabular}{l c c c} \hline  

Fluid \,&$\nu$ $({\rm N s m^{-2}})$ \,&$\rho$ $({\rm kg m^{-3}})$\,&$\Gamma$ $({\rm N m^{-1}})$\,\\ \hline \hline 

Water	    			&0.0010 	&997		&$0.07 \pm 0.003$ \\

Water-glycerol-1 		&0.0098 	&1076 	&$0.07 \pm 0.003$\\
 
Water-glycerol-2	 	&0.1190 	&1108 	&$0.07 \pm 0.003$\\

Glycerol 			&1.5  	&1126 	&$0.07 \pm 0.003$\\ 

Polybutene L-50   	&0.22		&844  	&$0.03 \pm 0.003$\\

Polybutene H-300  	&30		&892		&-----  \\

Polybutene H-300		&100 		&892		&-----  \\
\hline \end{tabular} \end{center} 

\caption{The fluids used in the experiments. $\nu$ is the viscosity, $\rho$ the density and $\Gamma$ is the surface tension of the fluid. All data corresponds to 25$^0$ C except in case of polybutene H-300, where the lower $\nu$ is obtained by heating the tank to 50$^0$ C. Water-glycerol-1 and water-glycerol-2 contain 60$\%$ and 88$\%$ glycerol by weight, respectively.}

\label{fluids} 

\end{table}
\end{multicols} 
\end{document}